\documentclass{article}
\usepackage[utf8]{inputenc}
\usepackage{graphicx}
\usepackage[portrait, margin=0.8in]{geometry}
\usepackage{hyperref}
\usepackage{authblk}
\usepackage{todonotes}
\usepackage{braket}
\usepackage{amsmath,amsthm,amssymb}
\usepackage[sorting = none, backend = bibtex, style=numeric-comp]{biblatex}

\title{Quantum Annealing vs. QAOA: 127 Qubit Higher-Order Ising Problems on NISQ Computers}

\author[1]{Elijah Pelofske\thanks{Email: epelofske@lanl.gov}}
\author[1]{Andreas Bärtschi\thanks{Email: baertschi@lanl.gov}}
\author[1]{Stephan Eidenbenz}
\affil[1]{CCS-3 Information Sciences, Los Alamos National Laboratory}

\date{\vspace{-6ex}}

\begin{document}
\maketitle

\begin{abstract}
Quantum annealing (QA) and Quantum Alternating Operator Ansatz (QAOA) are both heuristic quantum algorithms intended for sampling optimal solutions of combinatorial optimization problems. In this article we implement a rigorous direct comparison between QA on D-Wave hardware and QAOA on IBMQ hardware. These two quantum algorithms are also compared against classical simulated annealing. The studied problems are instances of a class of Ising models, with variable assignments of $+1$ or $-1$, that contain cubic $ZZZ$ interactions (higher order terms) and match both the native connectivity of the Pegasus topology D-Wave chips and the heavy hexagonal lattice of the IBMQ chips. The novel QAOA implementation on the heavy hexagonal lattice has a CNOT depth of $6$ per round and allows for usage of an entire heavy hexagonal lattice. Experimentally, QAOA is executed on an ensemble of randomly generated Ising instances with a grid search over $1$ and $2$ round angles using all 127 programmable superconducting transmon qubits of ibm\_washington. The error suppression technique digital dynamical decoupling is also tested on all QAOA circuits. QA is executed on the same Ising instances with the programmable superconducting flux qubit devices D-Wave Advantage\_system4.1 and Advantage\_system6.1 using modified annealing schedules with pauses. We find that QA outperforms QAOA on all problem instances. We also find that dynamical decoupling enables 2-round QAOA to marginally outperform 1-round QAOA, which is not the case without dynamical decoupling. 

\end{abstract}

\section{Introduction}
\label{section:introduction}

Quantum annealing (QA) in the transverse field Ising model is an analog computation technology which utilizes quantum fluctuations in order to search for ground state solutions of a problem Hamiltonian \cite{Kadowaki_1998, morita2008mathematical, das2008colloquium, Hauke_2020, Yarkoni_2022}. D-Wave quantum annealers are programmable hardware implementations of quantum annealing which use superconducting flux qubits \cite{PhysRevX.4.021041, king2022coherent}. 

Quantum Alternating Operator Ansatz (QAOA) is a hybrid quantum classical algorithm for sampling combinatorial optimization problems \cite{Hadfield_2019, 9259949, Wang_2020}, the quantum component of which can be instantiated with a programmable gate-based universal quantum computer. The Quantum Approximate Optimization Algorithm \cite{https://doi.org/10.48550/arxiv.1411.4028} was the first variational algorithm of this type, which was then generalized to the Quantum Alternating Operator Ansatz algorithm \cite{Hadfield_2019}. 

QAOA is effectively a Trotterization of the Quantum Adiabatic Algorithm, and is overall similar to Quantum Annealing. In particular both algorithms address combinatorial optimization problems. The exact characteristics of how both QA and QAOA will scale to large system sizes is currently not fully understood, in particular because quantum hardware is still in the NISQ era \cite{Lotshaw_2022, PhysRevX.8.031016, king2021scaling}. For example, there is evidence that QAOA may be more difficult for classical computers to simulate than quantum annealing, which could make it a viable candidate for quantum advantage \cite{https://doi.org/10.48550/arxiv.1602.07674}. Quantum annealing in particular has been experimentally evaluated against classical algorithms in order to determine for what problem types and under what settings quantum annealing could provide a scaling advantage over the next best state-of-the-art classical approaches \cite{PhysRevX.8.031016, king2021scaling, Mandr__2016, Boixo_2014, https://doi.org/10.48550/arxiv.2210.04291}. Generally these results are encouraging and show that quantum annealing can indeed sample certain problem types better than classical methods such as simulated annealing. There have been a number of studies that directly compare Quantum Annealing and QAOA for a number of different sampling tasks \cite{https://doi.org/10.48550/arxiv.2302.02278, Pelofske_2021, 10.1145/3425607, https://doi.org/10.48550/arxiv.1901.01903, https://doi.org/10.48550/arxiv.2301.00520}, however this paper presents, to the best of our knowledge, the largest direct comparison between Quantum Annealing and QAOA to date. There have been experimental QAOA implementations which used up to 40 qubits \cite{Pagano_2020}, 27 qubits \cite{Weidenfeller2022scalingofquantum}, and 23 qubits \cite{Harrigan_2021}. There have also been QAOA experiments which had circuit depth up to 159 \cite{niroula2022constrained} and 148 \cite{https://doi.org/10.48550/arxiv.2209.15024}.

The contributions of this article are as follows:

\begin{enumerate}
    \item We provide a direct comparison between QAOA and Quantum Annealing in terms of experiments on D-Wave and IBMQ hardware. This comparison uses a comparable parameter search space for QA and QAOA, uses no minor embedding for quantum annealing, and uses short depth QAOA circuits, thus providing a fair comparison of the two algorithms. A comparison of this problem size has not been performed before to the best of our knowledge. We show that QAOA is better than random sampling, and quantum annealing clearly outperforms QAOA. A comparison against the classical heuristic algorithm simulated annealing is also presented. 
    \item The QAOA algorithm we present is tailored for short depth circuit construction on the heavy hexagonal lattice (CNOT depth of $6$ per round), therefore allowing full usage of any heavy hexagonal topology quantum processor in the future. We use all $127$ qubits of the \texttt{ibm\_washington} chip in order to execute the largest QAOA circuit, in terms of qubits, to date. Each QAOA circuit uses thousands of gate operations, making these results one of the largest quantum computing experiments performed to date. 
    \item The Ising models that are used to compare quantum annealing and QAOA are specifically constructed to include higher order terms, specifically three variable (cubic) terms. QAOA can directly implement higher order terms, and quantum annealing requires order reduction using auxiliary variables to implement these higher order terms. This is the largest experimental demonstration of QAOA with higher order terms to date. 
    \item In order to mitigate errors when executing the QAOA circuits, we utilize digital dynamical decoupling. This is the largest usage of dynamical decoupling in terms of qubit system size to date, and the results show that digital dynamical decoupling improves performance for two round QAOA, suggesting that it will be useful for computations with large numbers of qubits in the noisy regime. 
\end{enumerate}

In Section \ref{section:methods} the QAOA and QA hardware implementations, and the simulated annealing implementation are detailed. Section \ref{section:results} details the experimental results and how the two quantum algorithms compare, including how simulated annealing compares. Section \ref{section:discussion} concludes with what the results indicate and future research directions. The figures in this article are generated using matplotlib \cite{thomas_a_caswell_2021_5194481, Hunter:2007}, and Qiskit \cite{Qiskit} in Python 3. Code, data, and additional figures are available in a public Github repository \footnote{\url{https://github.com/lanl/QAOA_vs_QA}}.

\section{Methods}
\label{section:methods}

The Ising models are defined in Section \ref{section:methods_problem_instances}. In Section \ref{section:methods_QAOA_implementation} the QAOA circuit algorithm and hardware parameters are defined. In Section \ref{section:methods_QA_implementation} the quantum annealing implementation is defined. Section \ref{section:methods_SA_implementation} defines the simulated annealing implementation. 

\begin{table*}[t!]
\begin{center}
\begin{tabular}{ |c|p{2.6cm}|c|p{1.5cm}|p{3.3cm}| }
 \hline
 Device name & Topology/chip name & Available qubits & Available couplers/ CNOTs & Computation type \\ 
 \hline
 \hline
 \texttt{Advantage\_system4.1} & Pegasus $P_{16}$ & 5627 & 40279 & QA \\ 
 \hline
 \texttt{Advantage\_system6.1} & Pegasus $P_{16}$ & 5616 & 40135 & QA \\ 
 \hline
 \texttt{ibm\_washington} & Eagle r1 \newline heavy-hexagonal & 127 & 142 & Universal gate-model \\ 
 \hline
\end{tabular}
\end{center}
\caption{NISQ hardware summary at the time the experiments were executed. The hardware yield (e.g., the number of available qubits or two qubit interactions) for all of these devices can be less than the logical lattice because of hardware defects, and can also change over time if device calibration changes. }
\label{table:hardware_summary}
\end{table*}

\subsection{Ising Model Problem Instances}
\label{section:methods_problem_instances}

The NISQ computers which are used in this comparison are detailed in Table \ref{table:hardware_summary}; the clear difference between the D-Wave quantum annealers and \texttt{ibm\_washington} is the number of qubits that are available. The additional qubits available on the quantum annealers will allow us to embed multiple problem instances onto the chips. The current IBMQ devices have a graph topology referred to as the heavy-hexagonal lattice \cite{PhysRevX.10.011022}. Therefore, for a direct QAOA and QA comparison we would want to be able to create QAOA circuits which match the logical heavy-hexagonal lattice and the quantum annealer graph topology of Pegasus. For this direct comparison we target D-Wave quantum annealers with Pegasus graph hardware \cite{zbinden2020embedding, dattani2019pegasus} connectivities. The two current D-Wave quantum annealers with Pegasus hardware graphs have chip id names \texttt{Advantage\_system6.1} and \texttt{Advantage\_system4.1}. The goal for this direct comparison is that ideally we want problems which can be instantiated on \emph{all} three of the devices in Table \ref{table:hardware_summary}. In particular, we want these implementations to not be unfairly costly in terms of implementation overhead. For example we do not want to introduce unnecessary qubit swapping in the QAOA circuit because that would introduce larger circuit depths which would introduce more decoherence in the computation. We also do not want to introduce unnecessary minor-embedding in the problems for quantum annealers. 

The other property of these problem instances that is of interest is an introduction of \emph{higher order terms}, specifically cubic $ZZZ$ interactions \cite{PhysRevA.61.012302} also referred to as multi-body interactions \cite{chancellor2017circuit}, in addition to random linear and quadratic terms. These higher order terms require both QAOA and QA to be handle these higher order variable interactions, which is an additional test on the capability of both algorithms. QAOA can naturally handle higher order terms \cite{9779809}. Implementing high order terms with QA requires introducing auxiliary variables in order to perform order reduction to get a problem structure that is comprised of only linear and quadratic terms, so that it can be implemented on the hardware, but whose optimal solutions match the optimal solutions of the original high order polynomial (for the non-auxiliary variables) \cite{Hauke_2020, VALIANTE2021108102, 5444874, Pelofske_2022_tensor, jiang2018quantum}. 

Taking each of these characteristics into account, we create a class of random problems which follow the native device connectivities in Table \ref{table:hardware_summary}. The problem instances we will be considering are Ising models defined on the hardware connectivity graph of the heavy hexagonal lattice of the device, which for these experiments will be \texttt{ibm\_washington}. For a variable assignment vector $z = (z_0, \ldots, z_{n-1}) \in \{ +1,-1\}^n$, the random Ising model is defined as

\begin{equation}
    C(z) = \sum_{v \in V} d_v \cdot z_v + \sum_{(i,j) \in E} d_{i,j} \cdot  z_i \cdot z_j + \sum_{l \in W} d_{l,n_1(l),n_2(l)} \cdot z_l \cdot z_{n_1(l)} \cdot z_{n_2(l)}
    \label{equation:problem_instance}
\end{equation}

Eq.~\eqref{equation:problem_instance} defines the class of random minimization Ising models with cubic terms as follows. Any heavy hexagonal lattice is a bipartite graph with vertices $V = \{0,\ldots,n-1\}$ partitioned as $V=V_2 \cup V_3$, where $V_3$ consists of vertices with a maximum degree of $3$, and $V_2$ consists of vertices with a maximum degree of $2$. $E \subset V_2 \times V_3$ is the edge set representing available two qubit gates (in this case CNOTs where we choose targets $i \in V_2$ and controls $j\in V_3$). $W$ is the set of vertices in $V_2$ that all have degree exactly equal to $2$. $n_1$ is a function that gives the qubit (variable) index of the first of the two neighbors of a degree-2 node and $n_2$ provides the qubit (variable) index of the second of the two neighbors of any degree-2 node. Thus $d_v$, $d_{i,j}$, and $d_{l,n_1(l),n_2(l)}$ are all coefficients representing the random selection of the linear, quadratic, and cubic coefficients, respectively. These coefficients could be drawn from any distribution - in this paper we draw the coefficients from $\{+1, -1\}$ with probability $0.5$. Eq.~\eqref{equation:problem_instance} therefore defines how to compute the objective function for a given variable assignment vector $z$.

\begin{figure}[t!]
    \centering
    \includegraphics[width=0.49\textwidth]{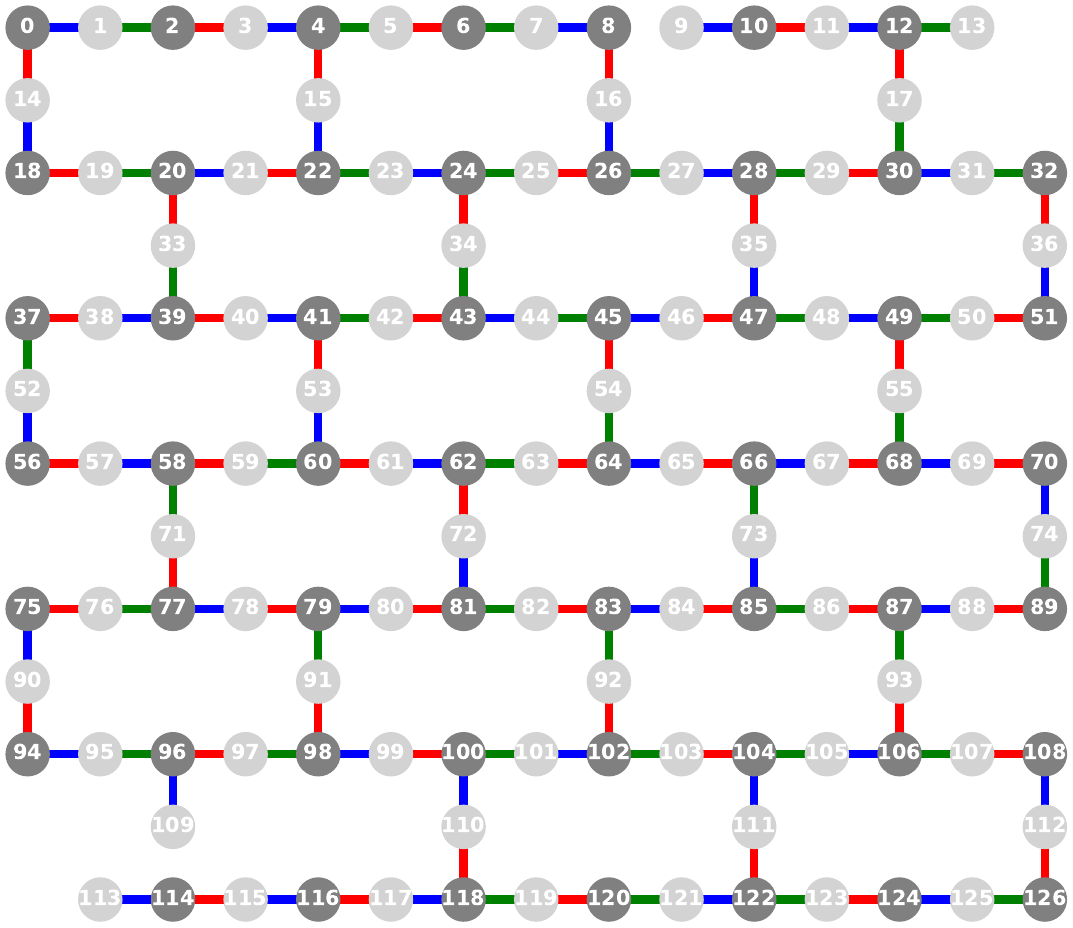}
    \includegraphics[width=0.49\textwidth]{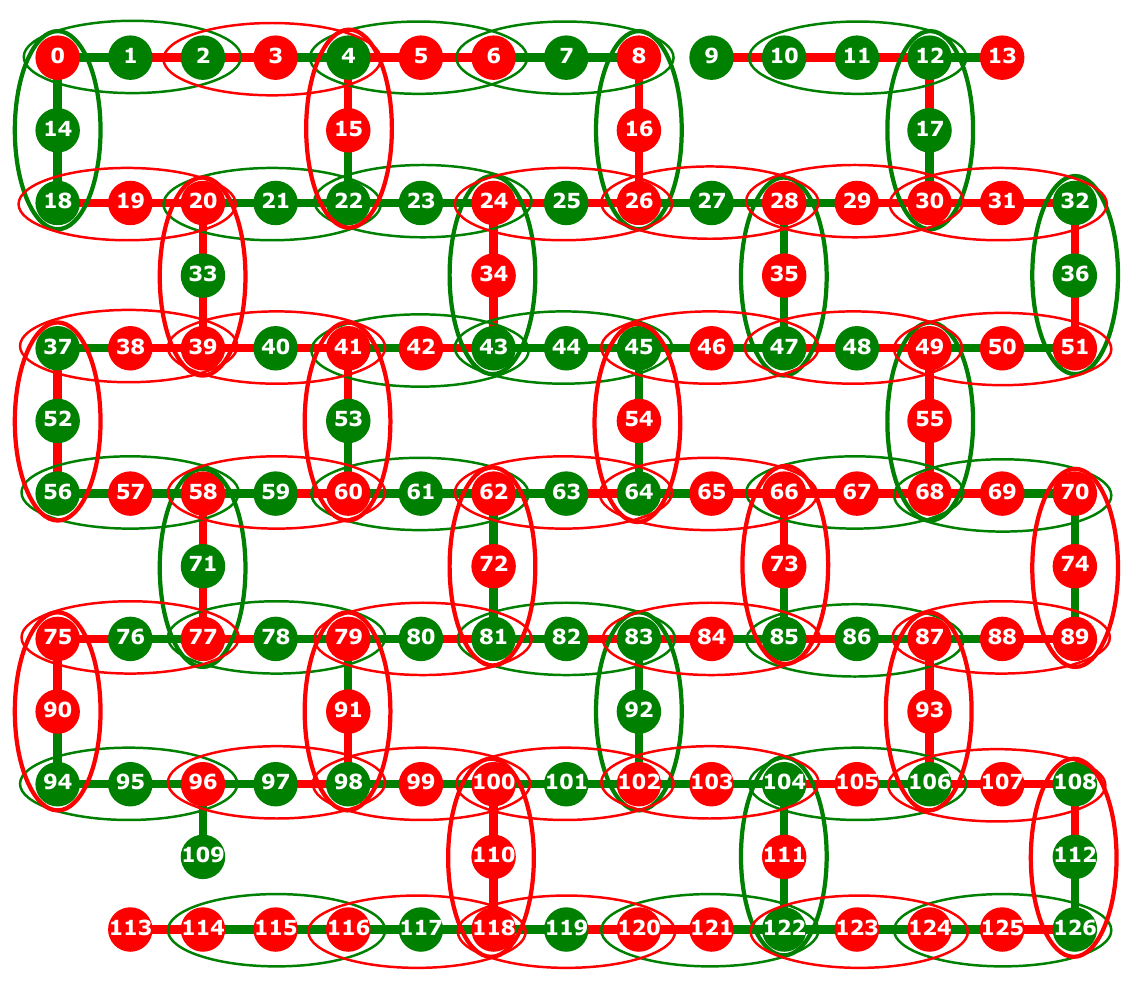}
    \caption{\textbf{Left:} \texttt{ibm\_washington} graph connectivity, where qubits are connected by CNOT (also referred to as \texttt{cx}) gates. The ideal lattice is called the heavy-hexagonal lattice. Note that there are two missing graph edges from the lattice between qubits 8-9 and 109-114. The total number of qubits (nodes) is $127$. The edges of the graph are three colored (red, blue, and green) such that no node shares two or more edges with the same color. The node colorings of light and dark gray show that the heavy hexagonal lattice is bipartite (meaning it can be partitioned into two disjoint sets). The three edge coloring is consistent with the QAOA circuit construction in Figure \ref{fig:QAOA_circuit_description}. \textbf{Right:} Example of a single random problem instance with cubic terms (see Eq.~\eqref{equation:problem_instance}) on the \texttt{ibm\_washington} graph. The linear and quadratic terms are shown using two distinct colors (red and green). The nodes and edges colored red denote a weight of $-1$ and the nodes and edges colored green denote a weight of $+1$. The cubic terms are represented by ovals around the three qubits which define the cubic variable interactions. Like the linear and quadratic terms, the color of the oval representing the cubic terms represents the sign of the weight on the terms, where green is $+1$ and red is $-1$. 
    }
    \label{fig:ibm_washington_graph}
\end{figure}

The heavy hexagonal topology of \texttt{ibm\_washington}, along with an overlay showing one of the random problem instances with cubic terms defined on \texttt{ibm\_washington}, is shown in Figure \ref{fig:ibm_washington_graph}. Each term coefficient was chosen to be either $+1$ or $-1$, which in part helps to mitigate the potential problem of limited precision for the programming control on all of the NISQ devices. $10$ random instances of this class of problems are generated and sampled using QAOA and QA, the implementations of each will be discussed next.

\subsection{Quantum Alternating Operator Ansatz}
\label{section:methods_QAOA_implementation}

Given a combinatorial optimization problem over inputs $z\in \{+1,-1\}^n$, let $C(z)\colon \{+1,-1\}^n \to \mathbb{R}$ be the objective function which evaluates the cost of the solution vector $z$. For a maximization (or minimization) problem, the goal is to find a variable assignment vector $z$ for which $f(z)$ is maximized (or minimized). The QAOA algorithm consists of the following components:

\begin{itemize}
    \item an initial state $\ket{\psi}$,
    \item a \texttt{phase separating} Cost Hamiltonian $H_C$,\\
    which is derived from $C(z)$ by replacing all spin variables $z_i$ by Pauli-Z operators $\sigma_i^z$
    \item a \texttt{mixing} Hamiltonian $H_M$; in our case, we use the standard transverse field mixer, which is the sum of the Pauli-X operators $\sigma_i^x$
    \item an integer $p\geq 1$, the number of rounds to run the algorithm,
    \item two real vectors $\vec{\gamma} = (\gamma_1,...,\gamma_p)$ and $\vec{\beta} = (\beta_1,...,\beta_p)$, each with length $p$.
\end{itemize}

The algorithm consists of preparing the initial state $\ket{\psi}$, then applying $p$ rounds of the alternating simulation of the phase separating Hamiltonian and the mixing Hamiltonian:

\begin{equation}
    \ket{\vec{\gamma},\vec{\beta}} = \underbrace{e^{-i\beta_p H_M} e^{-i\gamma_p H_P}}_{\text{round }p}\cdots \underbrace{e^{-i\beta_1 H_M} e^{-i\gamma_1 H_P}}_{\text{round }1} \ket{\psi}
\end{equation}

Within reach round, $H_P$ is applied first, which separates the basis states of the state vector by phases $e^{-i\gamma f(x)}$. $H_M$ then provides parameterized interference between solutions of different cost values. After $p$ rounds, the state $\ket{\vec{\gamma},\vec{\beta}}$ is measured in the computational basis and returns a sample solution $y$ of cost value $f(y)$ with probability $|\braket{y|\vec{\gamma},\vec{\beta}}|^2$. 

The aim of QAOA is to prepare the state $\ket{\vec{\gamma},\vec{\beta}}$ from which we can sample a solution $y$ with high cost value $f(y)$. Therefore, in order to use QAOA the task is to find angles $\vec{\gamma}$ and $\vec{\beta}$ such that the expectation value $\braket{\vec{\gamma},\vec{\beta}|H_P|\vec{\gamma},\vec{\beta}}$ is large ($-H_P$ for minimization problems). In the limit $p \rightarrow \infty$, QAOA is effectively a Trotterization of of the Quantum Adiabatic Algorithm, and in general as we increase $p$ we expect to see a corresponding increase in the probability of sampling the optimal solution \cite{https://doi.org/10.48550/arxiv.2202.00648}. The challenge is the classical outer loop component of finding the good angles $\vec{\gamma}$ and $\vec{\beta}$ for all rounds $p$, which has a high computational cost as $p$ increases.

Variational quantum algorithms, such as QAOA, have been a subject of large amount of attention, in large part because of the problem domains that variational algorithms can address (such as combinatorial optimization) \cite{Cerezo_2021}. One of the challenges however with variational quantum algorithms is that the classical component of parameter selection, in the case of QAOA this is the angle finding problem, is not solved and is even more difficult when noise is present in the computation \cite{wang2021noise}. Typically the optimal angles for QAOA are computed exactly for small problem instances \cite{Zhu_2023, Pelofske_2021}. However, in this case the angle finding approach we will use is a reasonably high resolution gridsearch over the possible angles. Note however that a fine gridsearch scales exponentially with the number of QAOA rounds $p$, and therefore is not advisable for practical high round QAOA \cite{https://doi.org/10.48550/arxiv.1411.4028, 9259949}. Exactly computing what the optimal angles are for problems of this size would be quite computationally intensive, especially with the introduction of higher order terms. We leave the problem of exactly computing the optimal QAOA angles to future work. 

\begin{figure}[t!]
    \centering
    \includegraphics[width=1.01\textwidth]{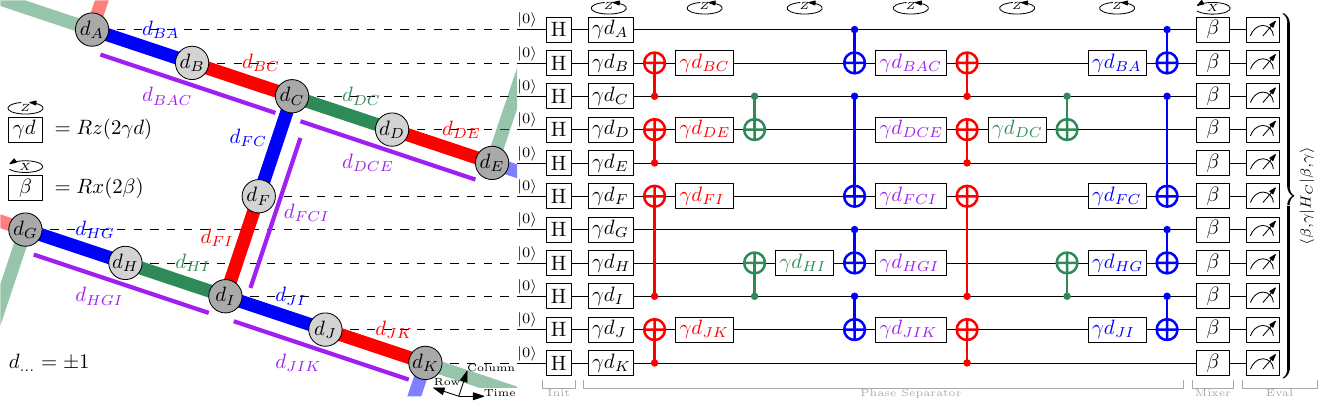}
    \caption{A 1-round QAOA circuit:
    \textbf{(left)} The problem instance is a hardware-native bipartite graph with an arbitrary 3-edge-coloring given by K\H{o}nig's line coloring theorem. 
    \textbf{(right)} Any quadratic term (colored edge) gives rise to a combination of two CNOTs and a Rz-rotation in the phase separator, giving a CNOT depth of 6 due to the degree-3 nodes. When targeting the degree-2 nodes with the CNOT gates, these constructions can be nested, leading to no overhead when implementing the three-qubit terms: these always have a degree-2 node in the middle (see Eq.~\eqref{equation:problem_instance}). 
    }
    \label{fig:QAOA_circuit_description}
\end{figure}

Figure \ref{fig:QAOA_circuit_description} describes the short depth QAOA circuit construction for sampling the higher order Ising test instance. This algorithm can be applied to any heavy hexagonal lattice topology, which allows for executing the QAOA circuits on the $127$ variable instances on the IBMQ \texttt{ibm\_washington} backend. For the class of Ising models with higher order terms defined in Section \ref{section:methods_problem_instances}, the QAOA angle ranges which are used are $\gamma_1, \ldots, \gamma_p \in [0, \pi)$ and $\beta_1, \ldots, \beta_{p-1} \in [0, \pi), \beta_{p} \in [0, \frac{\pi}{2})$ where $p$ is the number of QAOA rounds. Note that the halving of the angle search space for $\beta$ applies when $p=1$. For optimizing the angles using the naive grid search for $p=1$, $\beta_0$ is varied over $60$ linearly spaced angles $\in [0, \frac{\pi}{2}]$ and $\gamma_0$ is varied over $120$ linearly spaced angles $\in [0, \pi]$. For the high resolution gridsearch for $p=2$, $\beta_1$ is varied over $5$ linearly spaced angles $\in [0, \frac{\pi}{2}]$ and $\gamma_0$, $\gamma_1$, and $\beta_0$ are varied over $11$ linearly spaced angles $\in [0, \pi]$. Therefore, for $p=2$ the angle gridsearch uses $6655$ separate circuit executions (for each of the $10$ problem instances), and for $p=1$ the angle gridsearch uses $7200$ separate circuit executions. Each circuit execution used $10,000$ samples in order to compute a robust distribution for each angle combination.

In order to mitigate decoherence on idle qubits, digital dynamical decoupling (DDD) is also tested for all QAOA circuits. Dynamical Decoupling is an open loop quantum control technique error suppression technique for mitigating decoherence on idle qubits \cite{Niu_2022, RevModPhys.88.041001, PhysRevLett.82.2417, Ali_Ahmed_2013, LaRose_2022, Kim_2023}. Dynamical decoupling can be implemented with pulse level quantum control, and digital dynamical decoupling can be implemented simply with circuit level instructions of sequences of gates which are identities \cite{LaRose_2022}. Note that digital dynamical decoupling is an approximation of pulse level dynamical decoupling. Dynamical decoupling has been experimentally demonstrated for superconducting qubit quantum processors including IBMQ devices \cite{Niu_2022, https://doi.org/10.48550/arxiv.2207.03670, PhysRevLett.121.220502}. Dynamical decoupling in particular is applicable for QAOA circuits because they can be relatively sparse and therefore have idle qubits \cite{Niu_2022}. DDD does not always effective at consistently reducing errors during computation (for example because of other control errors present on the device \cite{Ali_Ahmed_2013, Niu_2022}), and therefore the raw QAOA circuits are compared against the QAOA circuits with DDD in the experiments section. In order to apply the DDD sequences to the OpenQASM \cite{https://doi.org/10.48550/arxiv.1707.03429} QAOA circuits, the \texttt{PadDynamicalDecoupling} \footnote{\url{https://qiskit.org/documentation/locale/bn_BN/stubs/qiskit.transpiler.passes.PadDynamicalDecoupling.html}} method from Qiskit \cite{Qiskit} is used, with the \texttt{pulse\_alignment} parameter set based on the \texttt{ibm\_washington} backend properties. The circuit scheduling algorithm that is used for inserting the digital dynamical decoupling sequences is ALAP, which schedules the stop time of instructions as late as possible \footnote{\url{https://qiskit.org/documentation/apidoc/transpiler_passes.html}}. There are other scheduling algorithms that could be applied which may increase the efficacy of dynamical decoupling. There are different DDD gate sequences that can be applied, including Y-Y or X-X sequences. Because the X Pauli gate is already a native gate of the IBMQ device, the X-X DDD sequence is used for simplicity. 

Note that the variable states for the optimization problems are either $-1$ or $+1$, but the circuit measurement states are either $0$ or $1$. Therefore once the measurements are made on the QAOA circuits, for each variable in each sample the variable state mapping of $0 \rightarrow 1$, $1 \rightarrow -1$ is performed. For circuit execution on the superconducting transom qubit \texttt{ibm\_washington}, circuits are batched into \emph{jobs} where each job is composed of a group of at most $250$ circuits - the maximum number of circuits for a job on \texttt{ibm\_washington} is currently $300$, but we use $250$ in order to reduce job errors related to the size of jobs. Grouping circuits into jobs is helpful for reducing the total amount of compute time required to prepare and measure each circuit. When submitting the circuits to the backend, they are all first locally transpiled via Qiskit \cite{Qiskit} with \texttt{optimization\_level=3}. This transpilation converts the gateset to the \texttt{ibm\_washington} native gateset, and the transpiler optimization attempts to simplify the circuit where possible. The QAOA circuit execution on \texttt{ibm\_washington} spanned a large amount of time, and therefore the backend versions were not consistent. The exact backend software versions were \texttt{1.3.7}, \texttt{1.3.8}, \texttt{1.3.13}, \texttt{1.3.15}, \texttt{1.3.17}.

\subsection{Quantum Annealing}
\label{section:methods_QA_implementation}

Quantum annealing is a proposed type of quantum computation which uses quantum fluctuations, such as quantum tunneling, in order to search for the ground state of a user programmed Hamiltonian. Quantum annealing, in the case of the transverse field Ising model implemented on D-Wave hardware, is explicitly described by the system given in Eq.~\eqref{equation:QA_Hamiltonian}. The state begins at time zero purely in the transverse Hamiltonian state $\sum_i \sigma^x_i$, and then over the course of the anneal (parameterized by the \emph{annealing time}) the user programmed Ising is applied according the function $B(s)$. Together, $A(s)$ and $B(s)$ define the anneal schedules of the annealing process, and $s$ is referred to as the \emph{anneal fraction}. The standard anneal schedule that is used is a linear interpolation between $s=0$ and $s=1$. 

\begin{equation}
    H = - \frac{A(s)}{2} \Big( \sum_i^n \sigma^x_i \Big) + \frac{B(s)} {2} \Big( H_{ising} \Big)
    \label{equation:QA_Hamiltonian}
\end{equation}

\begin{figure}[t!]
    \centering
    \includegraphics[width=0.38\textwidth]{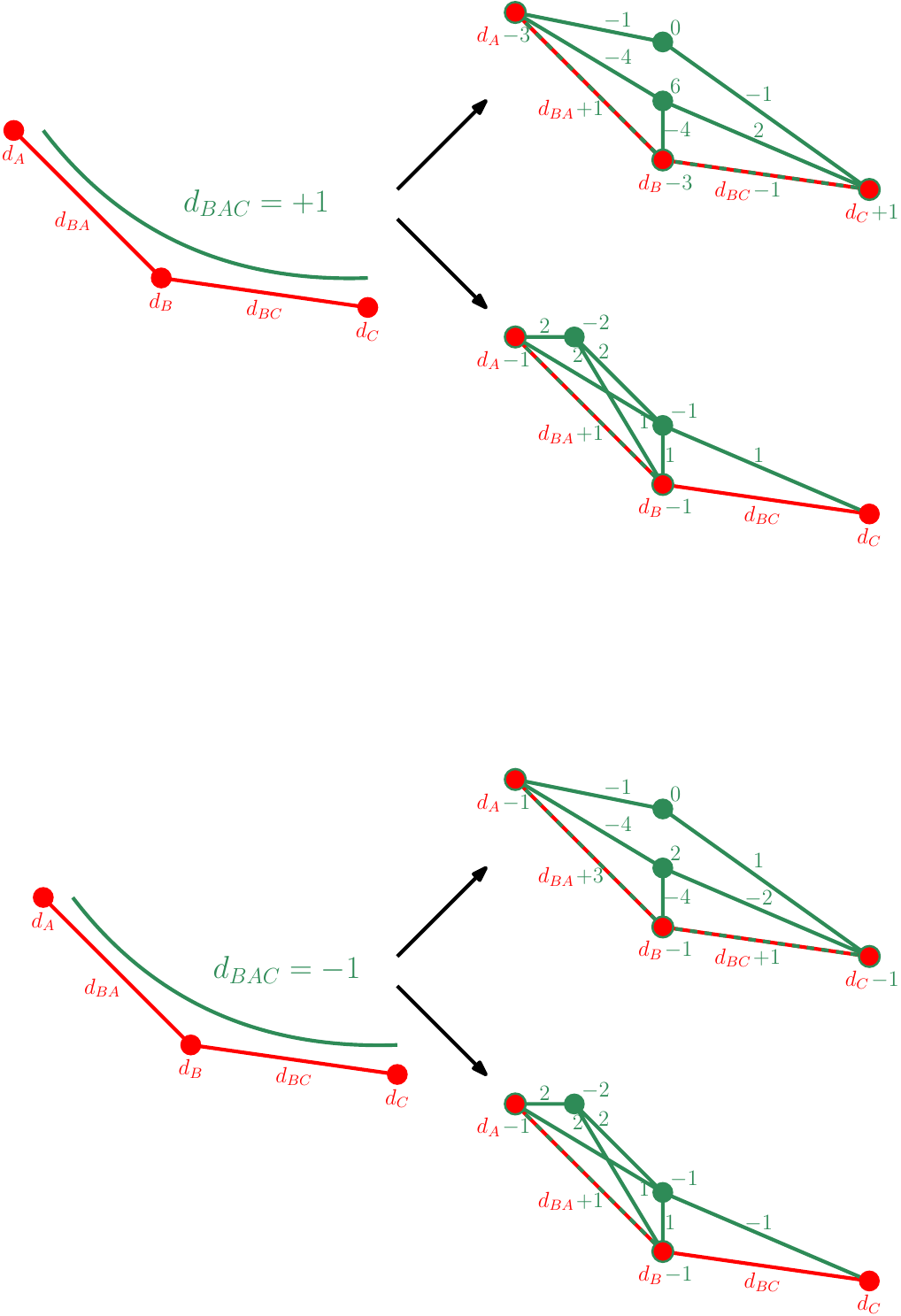}%
    \hfill
    \includegraphics[width=0.60\textwidth]{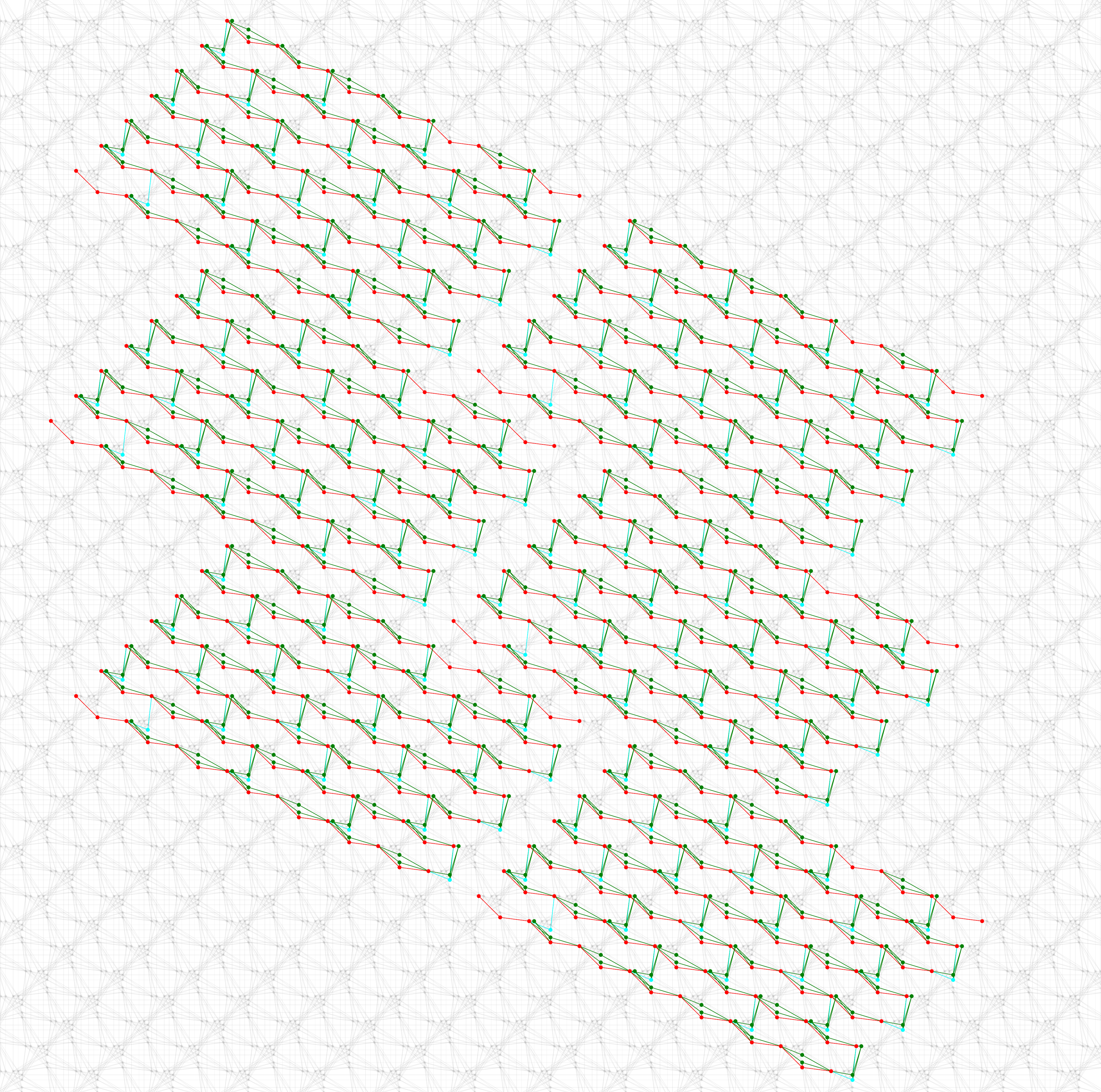}%
    \caption{
    \textbf{(left)} Two different embeddings for cubic $+1$/$-1$ terms. Each embedding needs two slack variable qubits. Our overall embedding alternates between these two cubic term embeddings. Any embedding with only one slack variable needs a 4-clique between the slack and the three original variables, which is not possible to embed for consecutive cubic terms.  \textbf{(right)} Embedding structures of the problem instances with higher order terms embedded in parallel (independently) $6$ times onto the logical Pegasus $P_{16}$ graph. The view of this graph has been slightly partitioned so that not all of the outer parts of the Pegasus chip are drawn. The light grey qubits and couplers indicate unused hardware regions. The cyan coloring on nodes and edges denote the vertical qubits and CNOTs on the \texttt{ibm\_washington} hardware graph (see Figure \ref{fig:ibm_washington_graph}). The red coloring on nodes and edges denote the horizontal lines of qubits and CNOTs on \texttt{ibm\_washington}. The green nodes and edges denote the order reduction auxiliary variables. Note that the top right hand and lower left hand qubits are not present on the \texttt{ibm\_washington} lattice - but for the purposes of generating the embeddings, these extra qubits are filled in to complete the lattice. }
    \label{fig:Logical_Pegasus_embeddings}
\end{figure}

\begin{figure}[t!]
    \centering
    \includegraphics[width=0.56\textwidth]{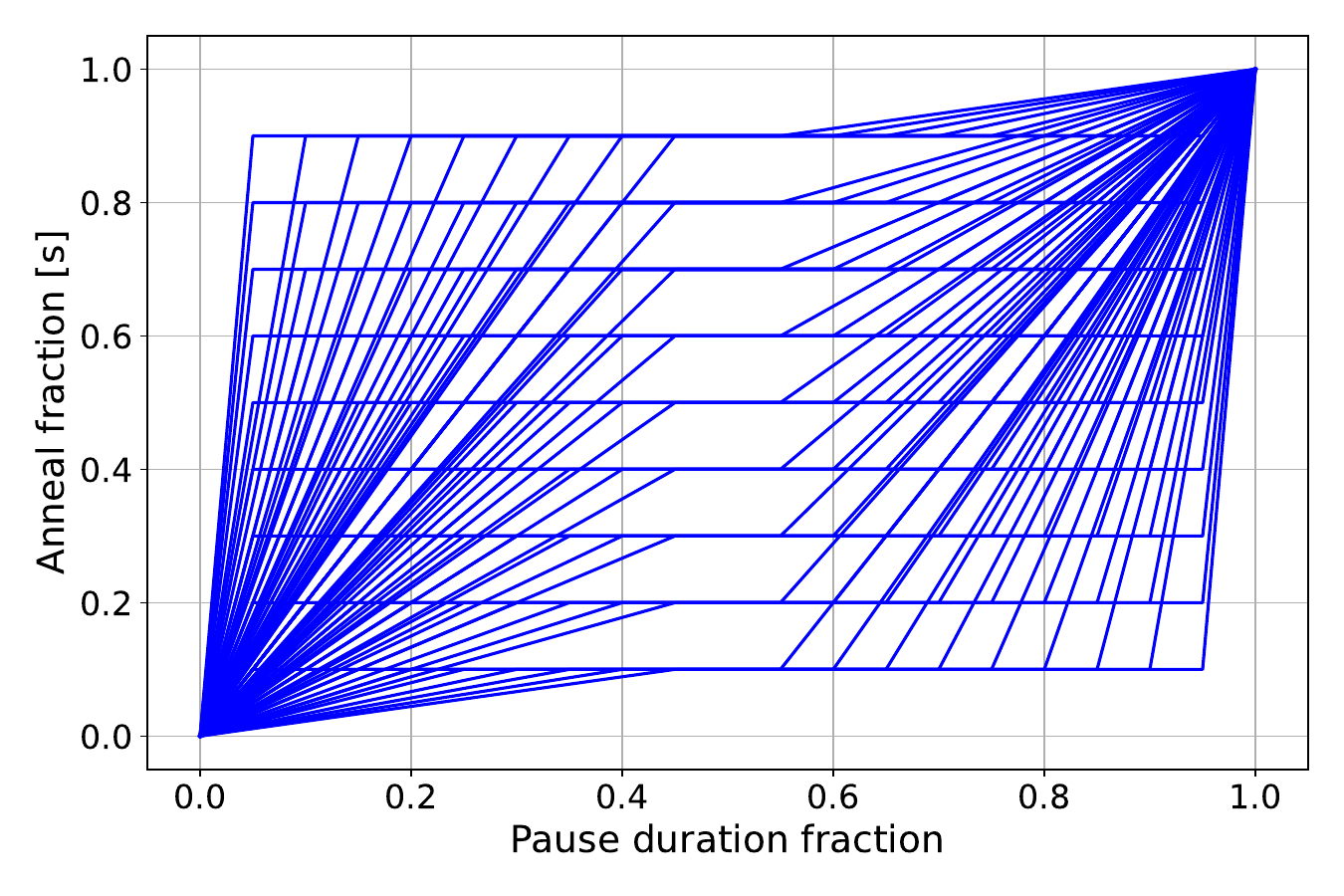}
    \caption{All modified (forward) quantum annealing schedules which are tested in order to find the best anneal schedule with a pause. The symmetric pause inserted into the normal linearly interpolated schedule defining the $A(s)$ and $B(s)$ functions can provide better ground state sampling probability. The anneal fraction at which this pause occurs is varied between $0.1$ and $0.9$ in steps of $0.1$. The pause duration, as a fraction of the total annealing time, is also varied between $0.1$ and $0.9$ in steps of $0.1$. Although not shown in this figure, the annealing times are also varied between $10$, $100$, $1000$, and $2000$ microseconds. }
    \label{fig:anneal_schedules}
\end{figure}

The adiabatic theorem states that if changes to the Hamiltonian of the system are sufficiently slow, the system will remain in the ground state of problem Hamiltonian, thereby providing a computational mechanism for computing the ground state of optimization problems. The user programmed Ising $H_{ising}$, acting on $n$ qubits, is defined in Eq.~\eqref{equation:problem_Hamiltonian}. The quadratic terms and the linear terms combined define the optimization problem instance that the annealing procedure will ideally find the ground state of. As with QAOA, the objective of quantum annealing is to find the variable assignment vector $z$ that minimizes the cost function which has the form of Eq.~\eqref{equation:problem_Hamiltonian}. 

\begin{equation}
    H_{ising} = \sum_{i}^n h_i \sigma_i^z + \sum_{i < j}^n J_{ij} \sigma_i^z \sigma_j^z
    \label{equation:problem_Hamiltonian}
\end{equation}

The goal is to be able to implement the Ising models defined in Section \ref{section:methods_problem_instances} on D-Wave quantum annealers. In order to implement the higher order terms, we will need to use order reduction in order to transform the cubic terms into linear and quadratic terms \cite{Hauke_2020, VALIANTE2021108102, 5444874, Pelofske_2022_tensor, jiang2018quantum}. This order reduction will result in using additional variables, usually called \emph{auxiliary} or \emph{slack} variables. Figure \ref{fig:Logical_Pegasus_embeddings} shows the embeddings of the problem instances onto the logical Pegasus $P_{16}$ graph, including the order reduction procedure which is used. The order reduction procedure outlined in Figure \ref{fig:Logical_Pegasus_embeddings} allows for direct embedding of the order reduced polynomials onto the hardware graph, regardless of whether the cubic term coefficient is $+1$ or $-1$. This order reduction ensures that the ground state(s) of the cubic term are also the ground states of the order reduced Ising. Additionally, this order reduction ensures that for every excited state of the cubic term, there are no slack variable assignments which result in the original variables having an energy less than or equal to the ground state of the original cubic term. This order reduction procedure allows any problem in the form of Eq.~\eqref{equation:problem_instance} to be mapped natively to quantum annealing hardware which accepts problems with the form of Eq.~\eqref{equation:problem_Hamiltonian}. Importantly, this procedure does not require minor-embedding, even including the auxiliary variables. 

In order to get more samples for the same QPU time, the other strategy that is employed is to embed multiple independent Ising model instances onto the hardware graph and thus be able to execute several instances in the same annealing cycle(s). This technique is referred to as \emph{parallel quantum annealing} \cite{Pelofske_2022, Pelofske_2022_tensor} or \emph{tiling} \footnote{\url{https://dwave-systemdocs.readthedocs.io/en/samplers/reference/composites/tiling.html}}. Figure \ref{fig:Logical_Pegasus_embeddings} (right) shows the parallel embeddings on a logical Pegasus graph. Because some of the logical embeddings may use a qubit or coupler which is missing on the actual hardware, less than $6$ parallel instances can be tiled onto the chips to be executed at the same time. For \texttt{Advantage\_system4.1}, $2$ independent embeddings of the problem instances could be created without encountering missing hardware. For \texttt{Advantage\_system6.1}, $3$ independent embeddings of the problem instances could be created. The structure of the heavy-hexagonal lattice onto Pegasus can be visually seen in Figure \ref{fig:Logical_Pegasus_embeddings}; the horizontal heavy-hex lines (Figure \ref{fig:ibm_washington_graph}) are mapped to diagonal Pegasus qubit lines that run from top left to bottom right of the square Pegasus graph rendering. Then the vertical heavy-hexagonal qubits are mapped to QA qubits in between the diagonal qubit lines. 

In order to optimize the quantum annealing parameters, with relatively similar complexity to the angle parameter search done for QAOA, the forward anneal schedule with pausing is optimized over a gridsearch. Pausing the anneal at the appropriate spot can provide higher chances of sampling the ground state \cite{PhysRevApplied.11.044083}. Figure \ref{fig:anneal_schedules} shows this anneal schedule search space - importantly the annealing times used in these schedule are also optimized for. The total number of QA parameters which are varied are $9$ anneal fractions, $9$ pause durations, and $4$ annealing times ($10$, $100$, $1000$, $2000$ microseconds). Therefore, the total number of parameter combinations which are considered in the grid search is $324$. $2000$ microseconds is the longest annealing time available on the current D-Wave quantum annealers. The number of anneals sampled for each D-Wave job was $500$. The annealing times and the anneal schedules were varied in a simple grid search. Readout and programming thermalization times are both set to $0$ microseconds. All other parameters are set to default, with the exception of the modified annealing schedule.

\subsection{Simulated Annealing implementation}
\label{section:methods_SA_implementation}

In order to provide a reasonable basis of comparison, the $10$ Ising model problem instances are also sampled using simulated annealing. Simulated annealing is a standard high accuracy and general purpose classical heuristic algorithm \cite{kirkpatrick1983optimization}, and has been used as a reasonable comparison against quantum algorithms \cite{PhysRevX.8.031016}. The simulated annealing implementation that we utilize is an open source implementation \footnote{\url{https://github.com/dwavesystems/dwave-neal}}. The settings we use are all set to default and $1000$ samples are drawn for each Ising model. The simulated annealing implementation does not natively handle higher order terms, and therefore order reduction must be applied to the Ising model's before being sampled by simulated annealing. Order reduction introduces additional variables into the computation. The order reduction is performed using the python package \emph{dimod} \footnote{\url{https://github.com/dwavesystems/dimod}}. The order reduction penalty \texttt{strength} is set to $2$, which ensures that the optimal solution of the original higher order Ising matches the order reduced Ising model (excluding the ancillary variables introduced by the order reduction). 

\begin{figure}[ht]
    \centering
    \includegraphics[width=0.89\textwidth]{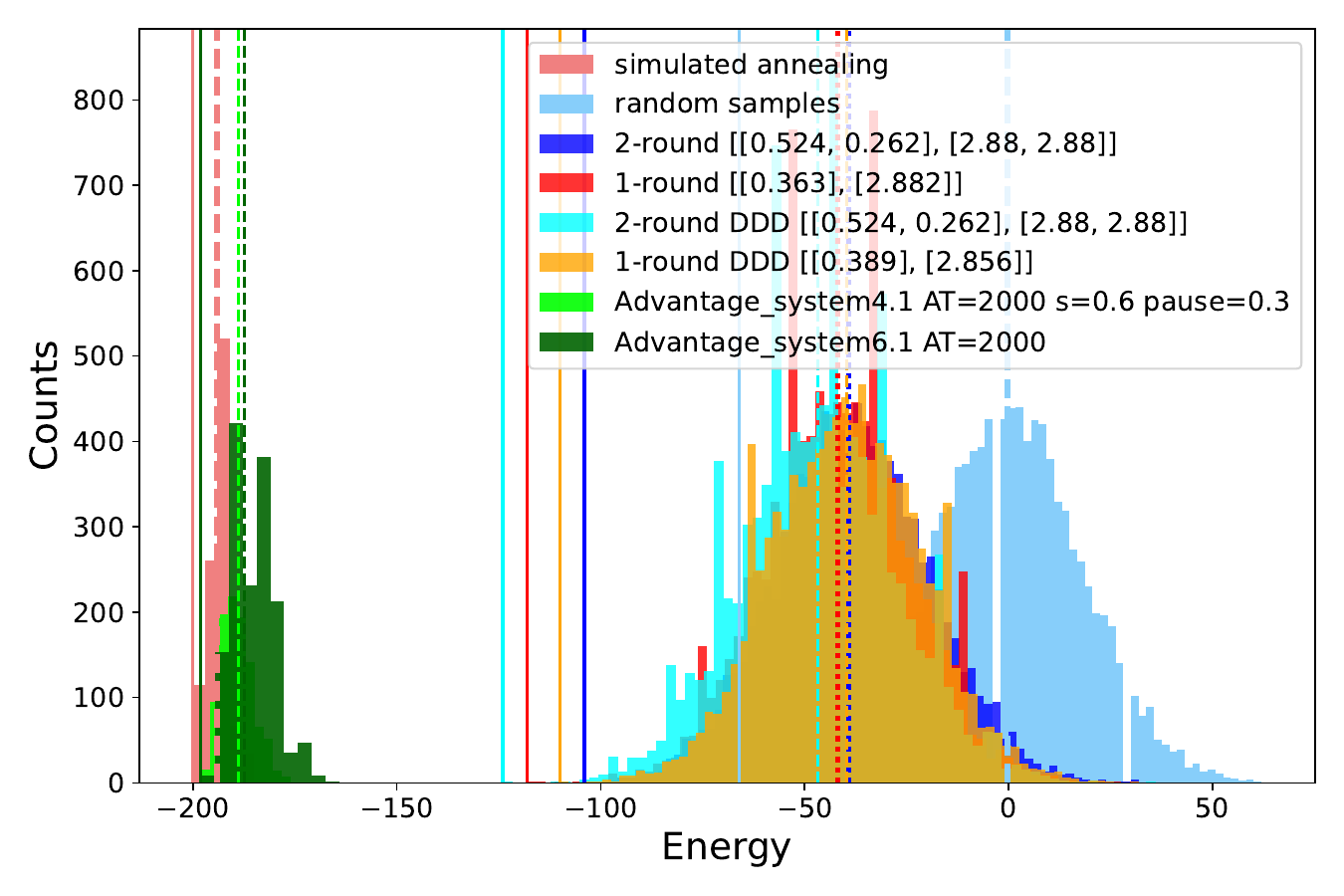}
    \caption{Direct objective function (e.g. energy) histogram comparison of QA and QAOA results for one of the $10$ minimization problem instances. A distribution of simulated annealing energies are also shown to provide a comparison against a reasonable classical heuristic. Here the energies being plotted are the full energy spectrum for the parameters which gave the minimum mean energy across the parameter grid searches performed across the QA and QAOA parameters. The optimal parameter combination for each distribution is given in the figure legend. For QA parameters, the annealing time in microseconds, the forward anneal schedule (symmetric) pause fraction, and anneal fraction, are given in the legend. If the default linearly interpolated quantum annealing schedule performed the best, only the annealing time reported in the legend. For the QAOA angle parameters, the format is $[\beta, \gamma]$, and are rounded to $3$ decimal places. The mean for each dataset is marked with vertical dashed lines and the minimum energy found in each dataset is marked with solid vertical lines. The energy histogram plots for the other 9 Ising models are shown in Figure \ref{fig:histogram_comparisons2}. }
    \label{fig:histogram_comparisons1}
\end{figure}

\section{Results}
\label{section:results}

\begin{figure}[t!]
    \centering
    \includegraphics[width=0.32\textwidth]{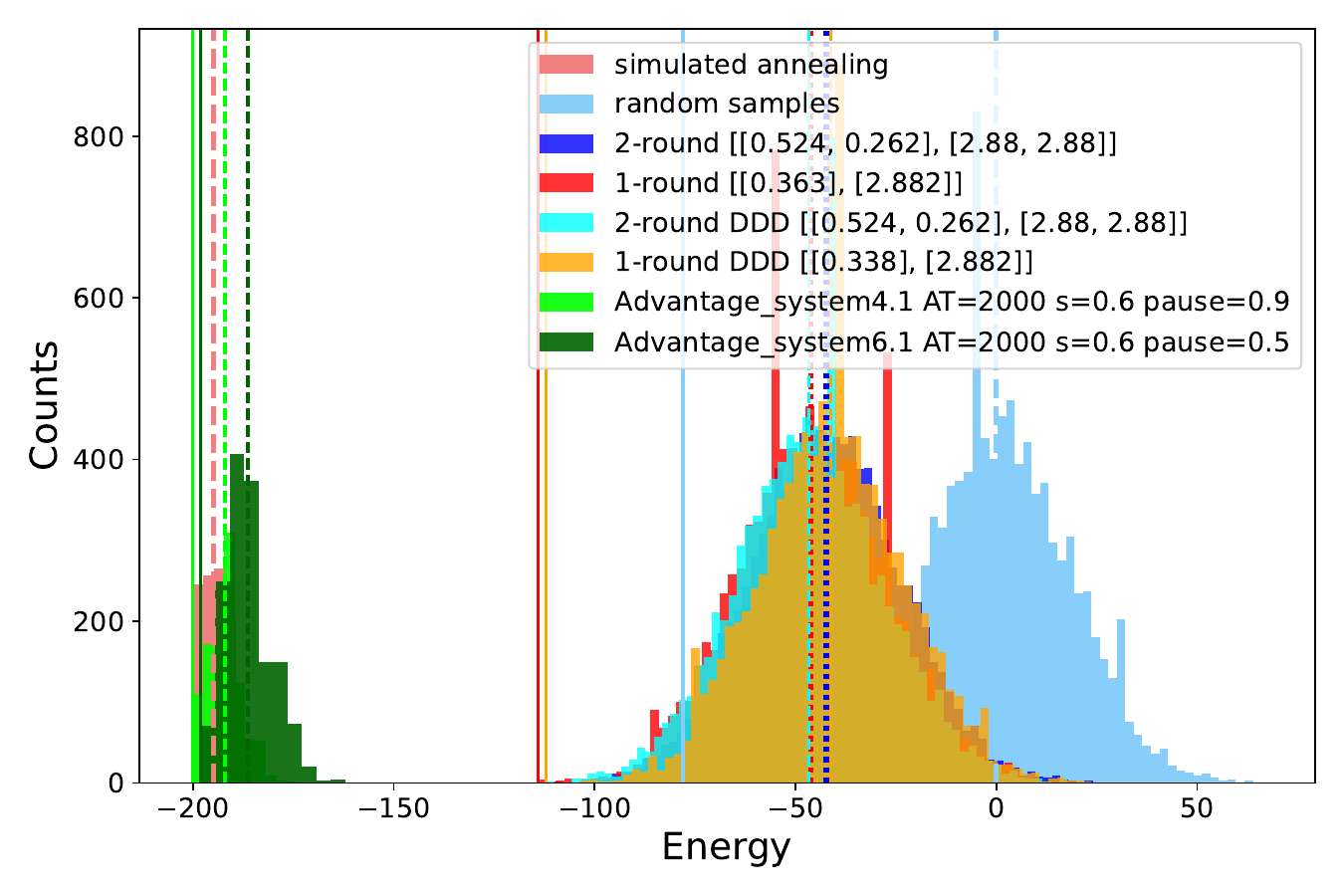}
    \includegraphics[width=0.32\textwidth]{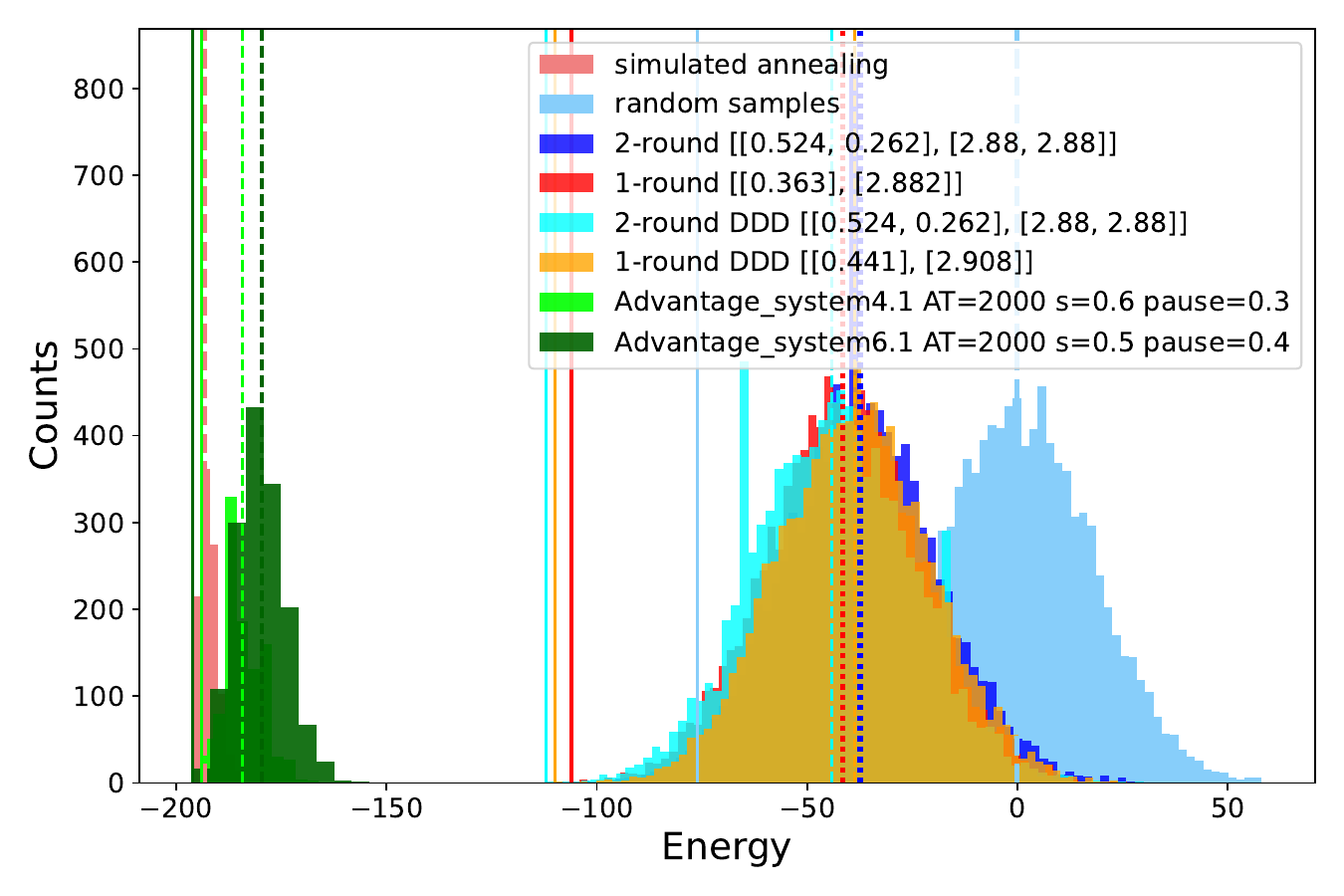}
    \includegraphics[width=0.32\textwidth]{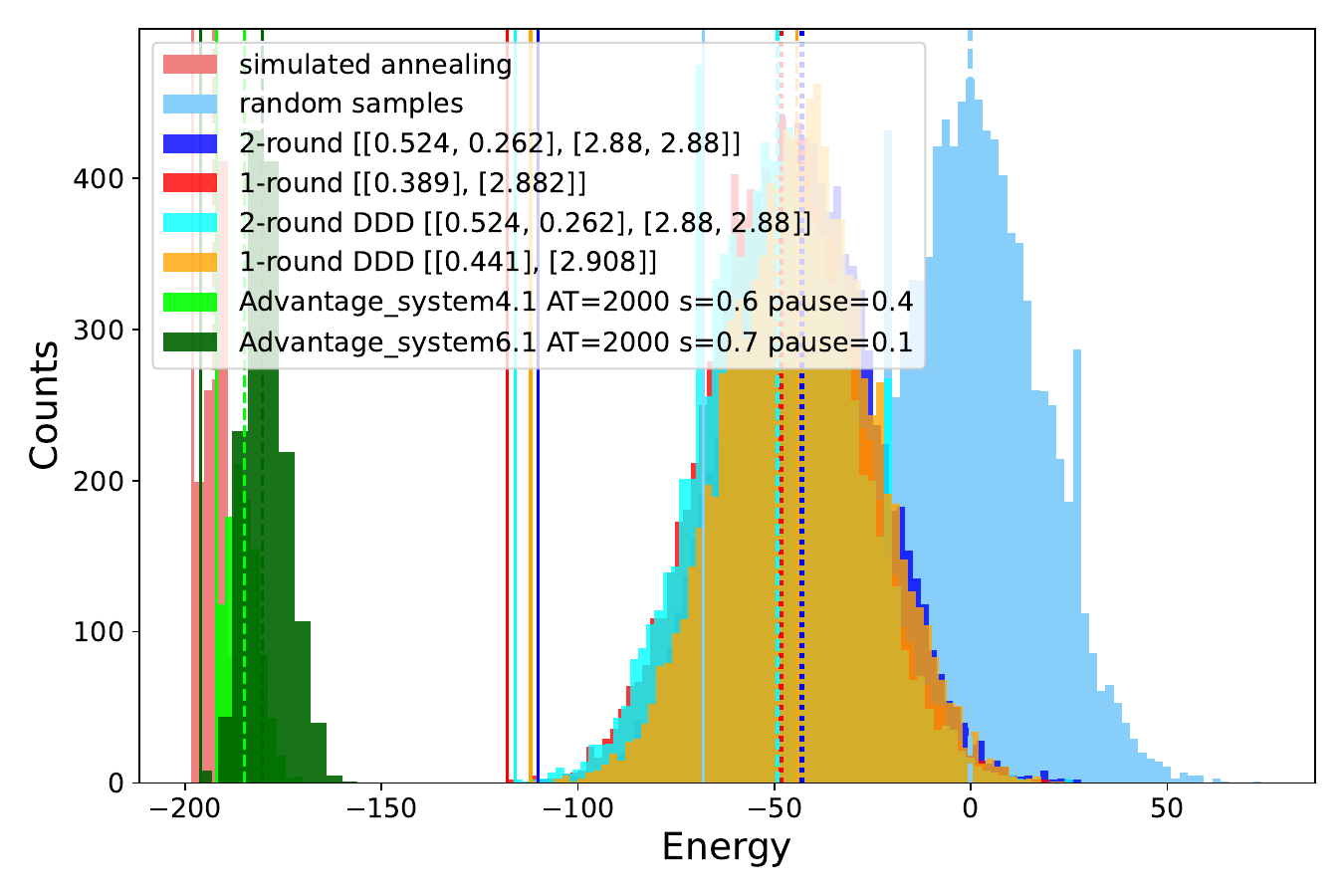}\\
    \includegraphics[width=0.32\textwidth]{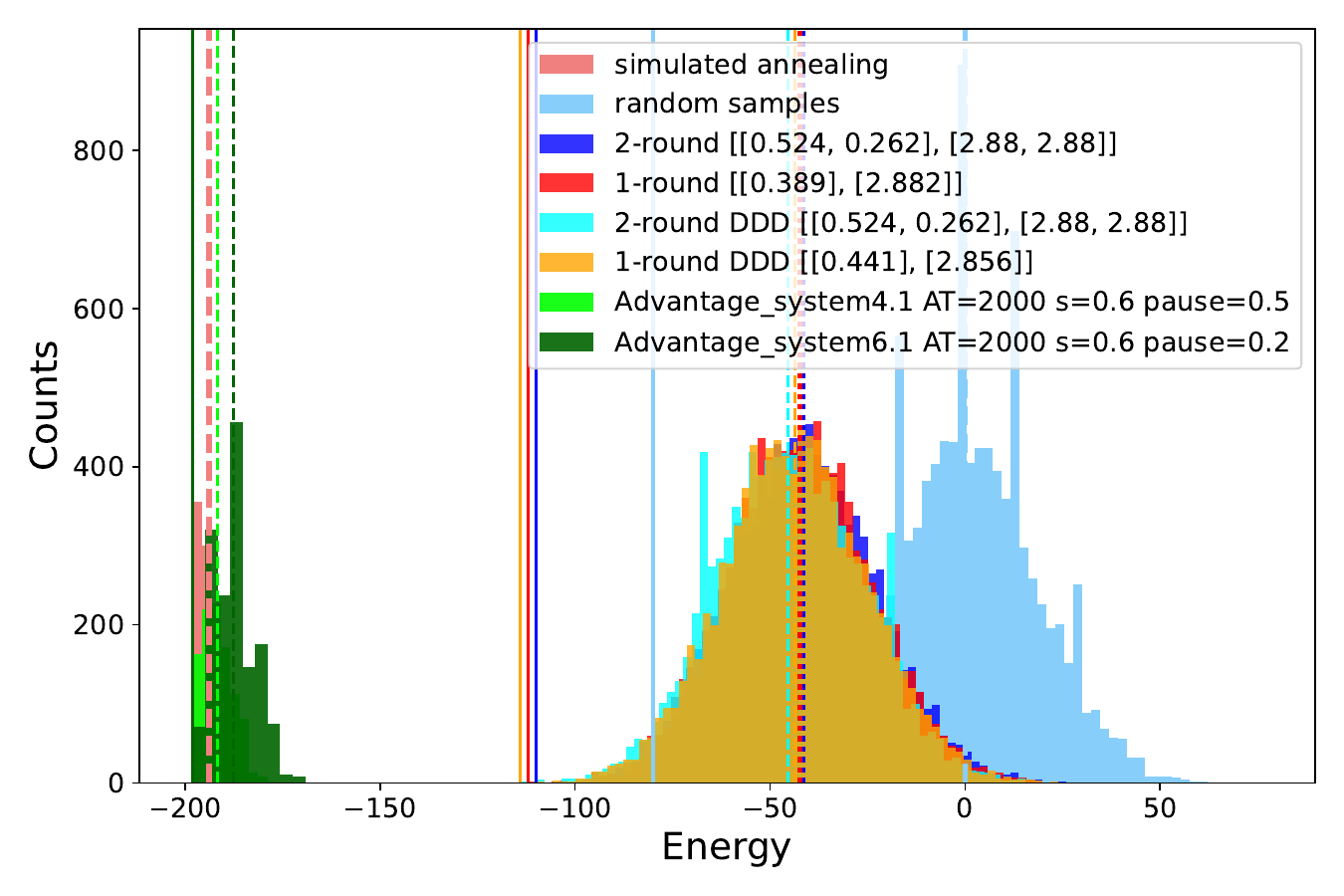}
    \includegraphics[width=0.32\textwidth]{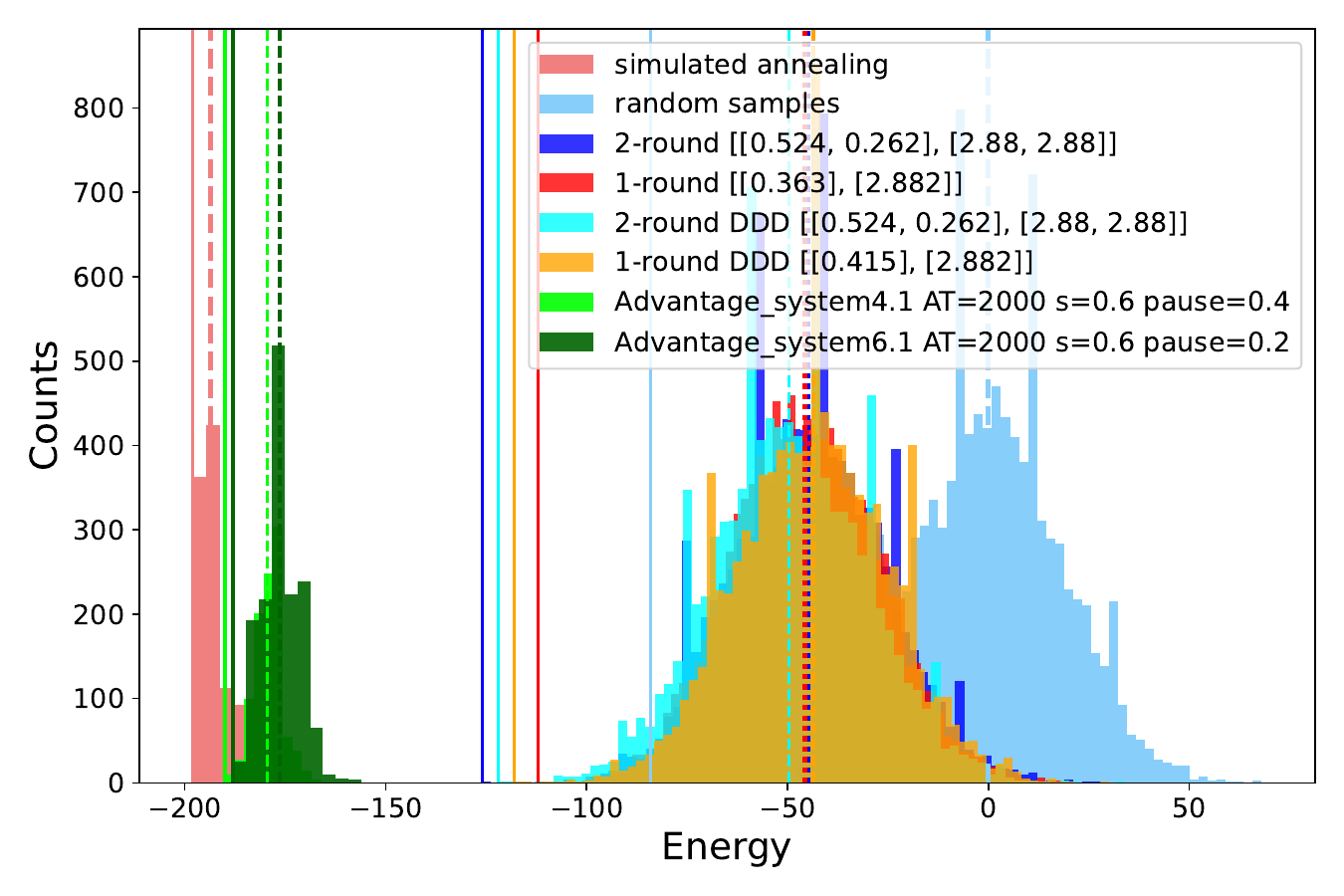}
    \includegraphics[width=0.32\textwidth]{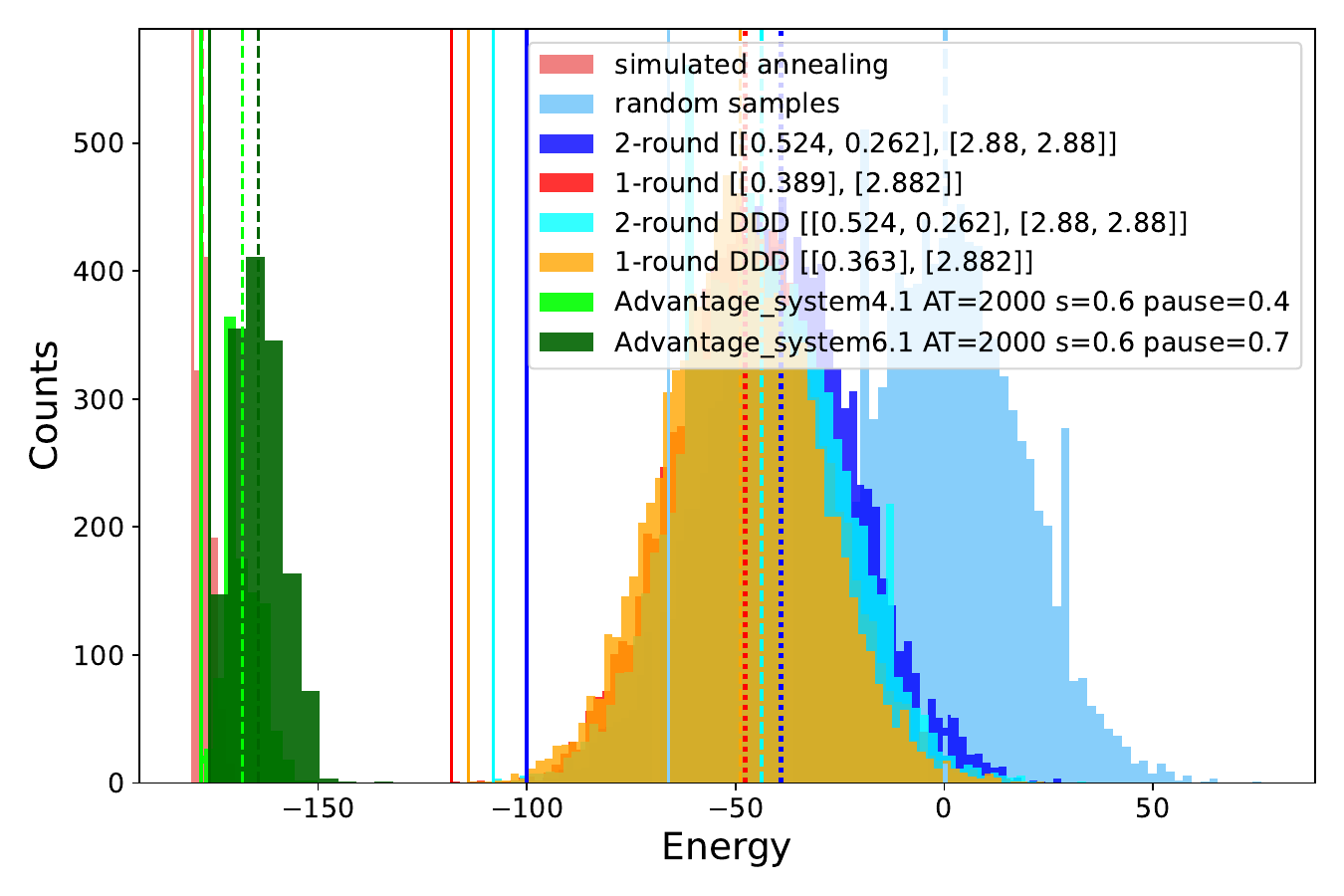}\\
    \includegraphics[width=0.32\textwidth]{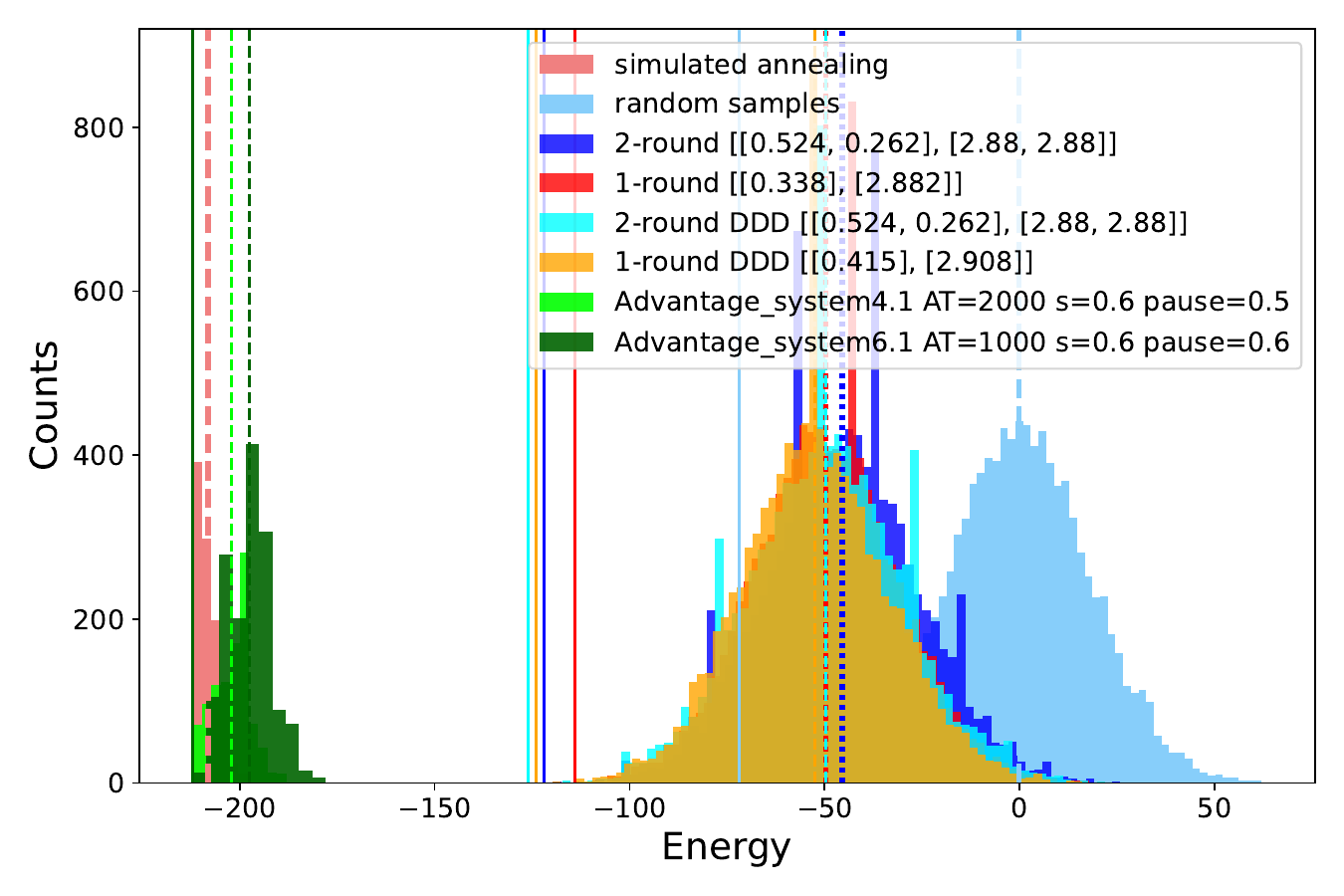}
    \includegraphics[width=0.32\textwidth]{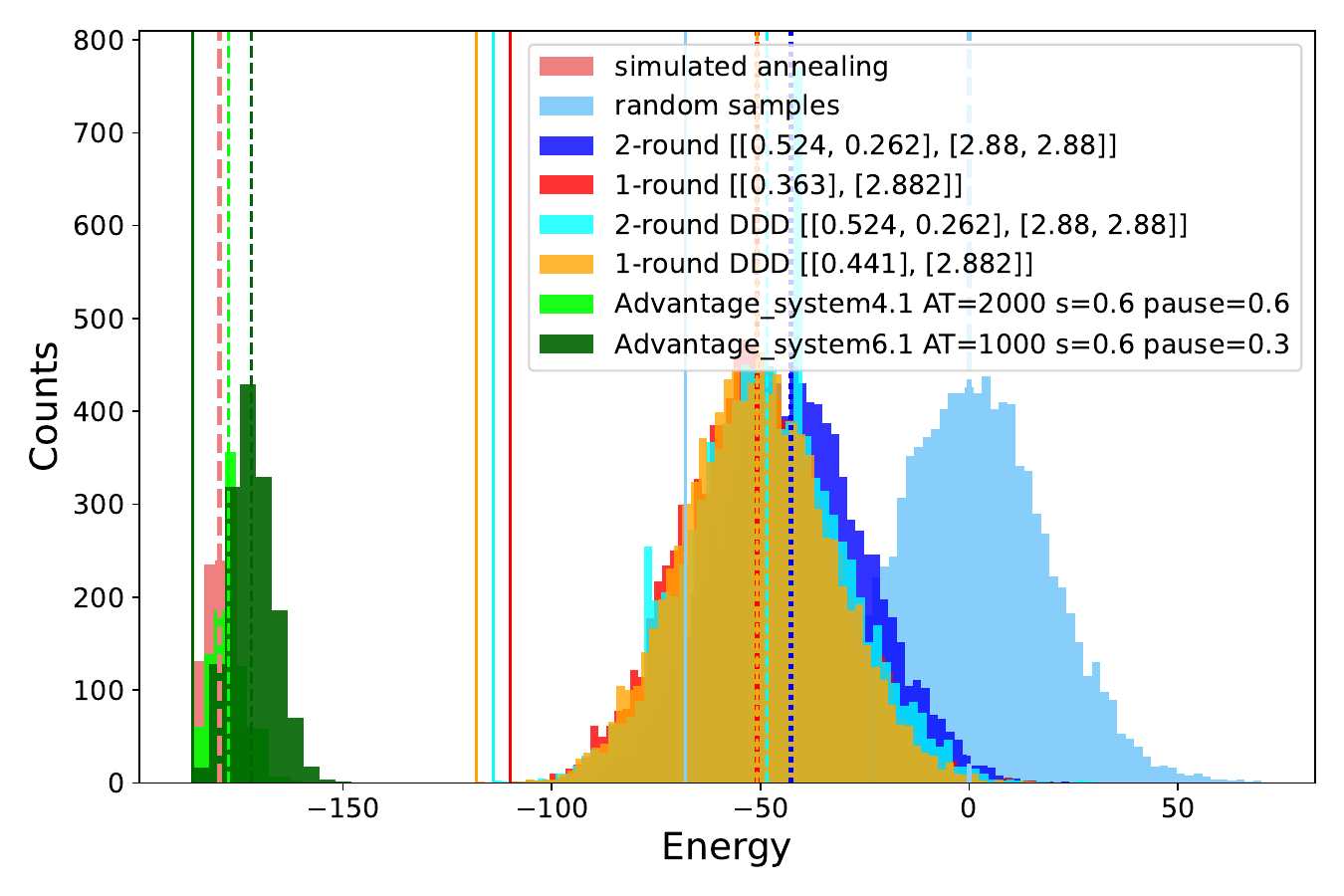}
    \includegraphics[width=0.32\textwidth]{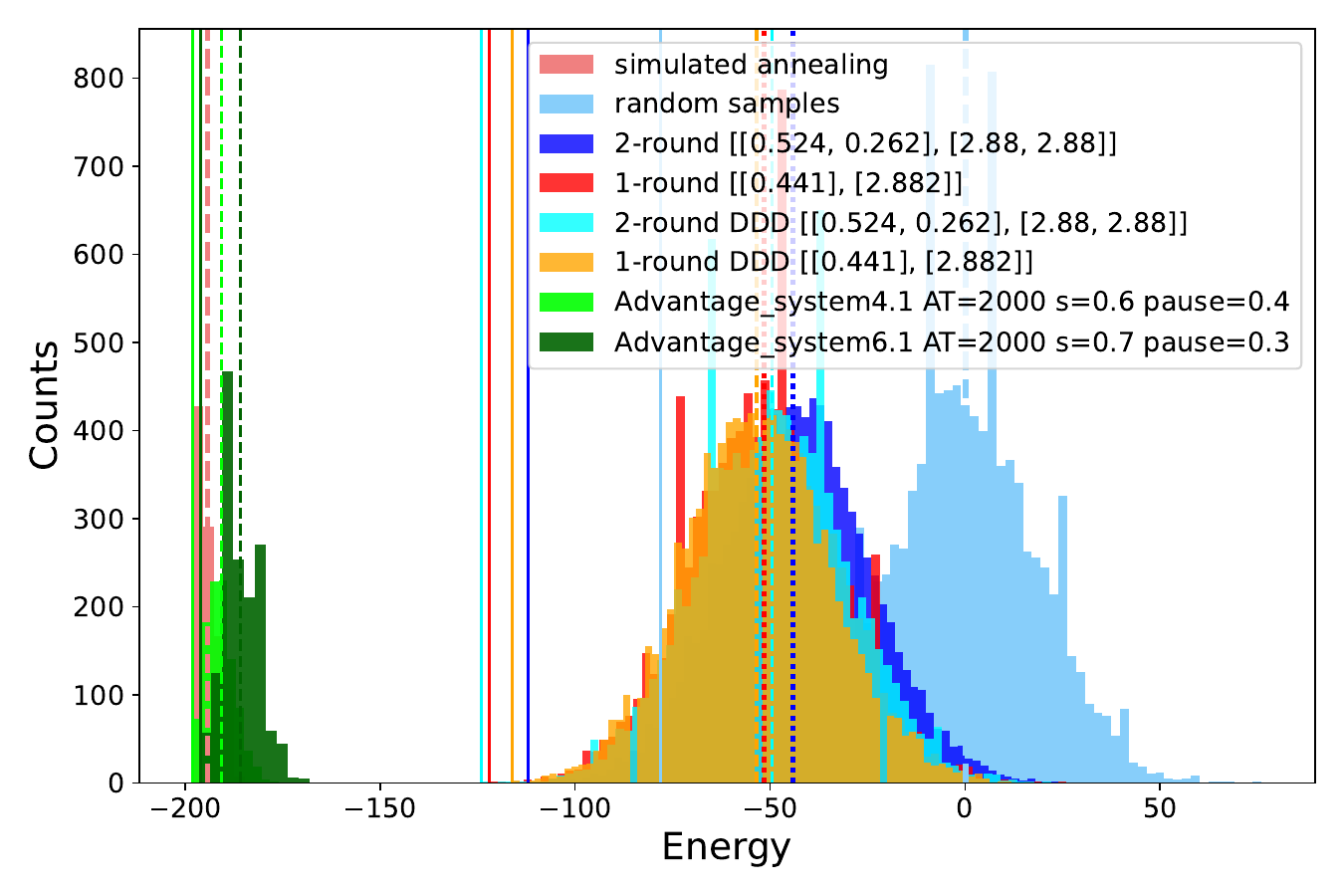}
    \caption{Direct energy histogram comparison of QA and QAOA results for the other nine problem instances, continuing from Figure \ref{fig:histogram_comparisons1}. The mean of each energy distribution is marked with vertical dashed lines, and the minimum energy of each dataset is marked with vertical solid lines. Note that for several of the distributions there are overlapping minimum energies. }
    \label{fig:histogram_comparisons2}
\end{figure}

Figures \ref{fig:histogram_comparisons1} and \ref{fig:histogram_comparisons2} combined show the detailed energy distributions for all $10$ cubic Ising models sampled using the best parameter choices found for QA and QAOA. These histograms include the four variants of QAOA - 1 and 2 rounds with and without digital dynamical decoupling. The histograms include $10000$ random samples (binomial distribution with $p=0.5$) on the $10$ Ising models.

\noindent
\textbf{QA performs better than QAOA:} The most notable observation across these histograms is that clearly quantum annealing results in better variable assignments compared to all tested variations of QAOA; this clear stratification of the algorithms capabilities is consistent across all $10$ problem instances. Notice that the minimum energies achieved by QAOA (marked by the solid vertical lines) do not reach the energy distribution sampled by the quantum annealers. The characteristics of each of the $10$ problem instances are slightly different, but this trend is very clear. 

\noindent
\textbf{QAOA performs better than random sampling:} Both QA and QAOA sampled better solutions than the $10000$ random samples. Although an obvious observation from the distributions in Figures \ref{fig:histogram_comparisons2} and \ref{fig:histogram_comparisons1}, it is not trivial that the QAOA samples had better objective function values compared to random sampling. The reason this is not trivial is because at sufficient circuit depth, which is not difficult to reach, the computation will entirely decohere and the computation will not be meaningful. This result is encouraging because it shows that short depth circuit constructions, combined with increasing scale of near term quantum computers, can begin to yield relevant computations for larger system sizes (in this case, $127$ variables). 

\noindent
\textbf{The effect of digital dynamical decoupling:} The dataset shown in Figure \ref{fig:histogram_comparisons2} also allows for a direct quantification of how successful the digital dynamical decoupling passes were at improving the QAOA circuit executions. Table \ref{table:QAOA_result_table_comparison} shows a comparison of the four QAOA implementations. For 2-round QAOA, DDD improved the mean sample energy for 10 out of the $10$ Ising models. For 1-round QAOA, DDD improved the mean sample energy for 4 out of the 10 problem instances. This shows that digital dynamical decoupling does not uniformly improve the performance of the QAOA circuits. This suggests that the qubits in the 2-round QAOA circuits have more available idle time compared to the 1-round QAOA circuits, which would allow for DDD to improve the circuit performance. The 2-round QAOA results had better average energy compared the 1-round results in 6 out of the 10 problem instances. 

\noindent
\textbf{Optimal parameter choices - QAOA:} The optimal 2-round QAOA angles for all $10$ problems with and without dynamical decoupling is the same. The optimal 1-round QAOA angles are not consistent across all problems, and even vary between the with and without DDD circuit executions. However, even though the exact optimal angle assignments are not consistent across all problems the, they are very close to each other which is notable because it indicates that the optimal angles may be identical or nearly identical but the search space is being obscured by the noise in the computation. 

\noindent
\textbf{Optimal parameter choices - QA:} Figure \ref{fig:histogram_comparisons2} also allows examination of how stable the different parameters are, both across the $10$ Ising models but also within each problem instance. In the case of quantum annealing, but the optimal annealing times are always $2000$ and the optimal pause schedule is not incredibly consistent with pause fraction durations ranging from $0.1$ to $0.9$ and with anneal fractions $s$ ranging from $0.5$ to $0.7$. 

\noindent
\textbf{D-Wave devices performance differences:} One last observation from Figure \ref{fig:histogram_comparisons2} is that there a small but consistent performance difference between the two quantum annealers; the slightly older generation \texttt{Advantage\_system4.1} yields lower mean energy than \texttt{Advantage\_system6.1}. Simulated annealing is comparable to the quantum annealing distributions, with simulated annealing performing marginally better than the quantum annealing distributions.

\begin{table*}[t!]
\begin{center}
\begin{tabular}{ |p{4.9cm}|p{0.9cm}|p{0.9cm}|p{2.5cm}|p{2.5cm}| } 
 \hline
  & $p=1$ & $p=2$ & $p=1$ with DDD & $p=2$ with DDD \\ 
 \hline
 \hline
 $p=1$ (no DDD) better than - & - & $10/10$ & $5/10$ & $4/10$ \\ 
 \hline
 $p=2$ (no DDD) better than - & $0/10$ & - & $2/10$ & $0/10$ \\ 
 \hline
 $p=1$ (with DDD) better than - & $5/10$ & $8/10$ & - & $4/10$ \\ 
 \hline
 $p=2$ (with DDD) better than - & $6/10$ & $10/10$ & $6/10$ & - \\ 
 \hline
\end{tabular}
\end{center}
\caption{How the four different QAOA implementations, one and two rounds with and without DDD, compare against each other in terms of in how many of the $10$ random instances each method was better than the other three methods in terms of mean objective function value across the $10000$ samples (for the best angle combination). There is a clear finding in the order of performance of the four methods; $p=2$ with no DDD performed the worse, $p=1$ with no DDD performed the next best, $p=1$ with DDD performed the next best, and $p=2$ with digital dynamical decoupling performed the best overall. }
\label{table:QAOA_result_table_comparison}
\end{table*}

\vspace{-0.3cm}

\section{Discussion}
\label{section:discussion}
It is of considerable interest to determine how effective quantum annealing and QAOA are at computing the optimal solutions of combinatorial optimization problems. Combinatorial optimization problems have wide reaching applicability, and being able to solve them faster or to get better heuristic solutions is a very relevant topic in computing. In this article, we have presented experimental results for a fair direct comparison of QAOA and quantum annealing, implemented on the state-of-the-art currently accessible quantum hardware via cloud computing. We leave more detailed benchmarking against state of the art classical solvers on these Ising model instances to future work. This research has specifically found the following:

\begin{enumerate}
    \item Quantum annealing finds higher quality solutions to the random test Ising models with higher order terms compared to the short depth QAOA $p=1$ and $p=2$ circuits, with reasonably fine grid searches over the QAOA angles and quantum annealing schedules with pauses. 
    \item QAOA performs noticeably better than random sampling - this is mostly due to the short depth QAOA circuit constructions which allow reasonably robust computations to be executed without the qubits decohering on current quantum computers. 
    \item The short depth QAOA circuit construction is notable because it allows for higher order terms in the Ising, and is scalable to a heavy-hexagonal lattice of any size, therefore this circuit construction can be used for future implementations of QAOA on devices with heavy-hexagonal lattices for heavy-hex native Ising models. 
    \item Dynamical decoupling can improve the computation of QAOA on NISQ computers. 
\end{enumerate}

\vspace{-0.5cm}

\section{Acknowledgments}
\label{section:acknowledgments}
This work was supported by the U.S. Department of Energy through the Los Alamos National Laboratory. Los Alamos National Laboratory is operated by Triad National Security, LLC, for the National Nuclear Security Administration of U.S. Department of Energy (Contract No. 89233218CNA000001).
The research presented in this article was supported by the Laboratory Directed Research and Development program of Los Alamos National Laboratory under project number 20220656ER and the NNSA's Advanced Simulation and Computing Beyond Moore's Law Program at Los Alamos National Laboratory.
This research used resources provided by the Darwin testbed at Los Alamos National Laboratory (LANL) which is funded by the Computational Systems and Software Environments subprogram of LANL's Advanced Simulation and Computing program (NNSA/DOE). 
This research used resources provided by the Los Alamos National Laboratory Institutional Computing Program.
We acknowledge the use of IBM Quantum services for this work. The views expressed are those of the authors, and do not reflect the official policy or position of IBM or the IBM Quantum team.
The authors would like to thank the anonymous reviewers for their helpful comments which helped to improve the manuscript. 

\noindent
LA-UR-22-33077

\printbibliography

\end{document}